# Gate-controlled supercurrent effect in dry-etched Dayem bridges of non-centrosymmetric niobium rhenium

J. Koch[1], C. Cirillo[2], S. Battisti[3], L. Ruf[1], Z. Makhdoumi Kakhaki[4], A. Paghi[3], A. Gulian[4], S. Teknowijoyo[4], G. De Simoni[3], F. Giazotto[3], C. Attanasio[2,5], E. Scheer[1†], A. Di Bernardo[1,5*]

1. *Department of Physics, University of Konstanz, Universitätsstraße 10, 78457 Konstanz, Germany.*
2. *CNR-Spin, c/o Università degli Studi di Salerno, via Giovanni Paolo II 132, I-84084, Fisciano (SA), Italy*
3. *NEST, Istituto Nanoscienze and Scuola Normale Superiore, Piazza San Silvestro 12, 56127 Pisa, Italy.*
4. *Advanced Physics Laboratory, Institute for Quantum Studies, Chapman University, Burtonsville, MD 20866, USA*
5. *Dipartimento di Fisica "E. R. Caianiello", Università degli Studi di Salerno, via Giovanni Paolo II 132, I-84084, Fisciano (SA), Italy.*

[†]Email: elke.scheer@uni-konstanz.de.
[*]Email: adibernardo@unisa.it.

**Abstract:**
The application of a gate voltage to control the superconducting current flowing through a nanoscale superconducting constriction, named as gate-controlled supercurrent (GCS), has raised great interest for fundamental and technological reasons. To gain a deeper understanding of this effect and develop superconducting technologies based on it, the material and physical parameters crucial for GCS must be identified. Top-down fabrication protocols should be also optimized to increase device scalability, although studies suggest that top-down fabricated devices are more resilient to show GCS. Here, we investigate gated superconducting nanobridges made with a top-down fabrication process from thin films of the non-centrosymmetric superconductor NbRe. Unlike other devices previously reported, our NbRe devices systematically exhibit GCS, when made in specific conditions, which paves the way for higher device scalability. Our results also suggest that surface properties of NbRe nanobridges and their modification during fabrication are key for GCS.

The recent discovery[1] that the superconducting critical current ($I_c$) of a nanoconstriction made from a superconductor (S) can be controlled via a gate voltage ($V_G$) has raised great interest for fundamental and technological reasons. These reasons have motivated studies on a variety of gated superconducting devices, made with different Ss, geometries and fabrication processes,[1]-[24] to determine under which conditions an applied $V_G$ can switch a S nanoconstriction from a superconducting state (with $I_c \neq 0$) to a resistive state (with $I_c = 0$). Although a $V_G$-driven modulation of the $I_c$, currently referred to as gate-controlled supercurrent (GCS),[22][25] has been observed in these studies,[1]-[24] the mechanism underlying GCS as well as the microscopic parameters and physical properties crucial for its observation remain not fully understood.[25]

To date, several mechanisms have been proposed to explain GCS which have been recently classified into four categories.[25] These categories include Fowler-Nordheim tunnelling of electrons[26] between the gate and S nanoconstriction across vacuum (scenario 1),[13][15] heating induced by phonons triggered by the leakage current ($I_{leak}$) flowing from the gate into the S nanoconstriction (scenario 2),[13],[16],[21] $I_{leak}$-induced phase fluctuations but without sizable heating (scenario 3),[16],[17],[20],[22]-[23] and microscopic mechanisms driven by the electric field associated to $V_G$ (scenario 4).[1]-[12],[18],[23]-[24],[27]-[30] Although some of these mechanisms (e.g., scenarios 1 and 2) may be at play in specific devices only (e.g., devices made on non-insulating substrates like Si),[25] no conclusive experiment has been reported that rules out one of the other two mechanisms (scenarios 3 and 4) and/or exactly quantifies their relative contributions towards GCS.

Understanding the mechanisms behind GCS is not only a fundamental challenge, but it may prove crucial also to enhance the performance of GCS-based devices. Figures of merit include the operational speed and the $V_G$ needed for a full $I_c$ suppression ($V_{G,offset}$). While speed may be limited in case of considerable heat dissipation (scenarios 1 and 2), GCS devices based on mechanism 3 or 4 may compete, in terms of speed, with existing superconducting logics.[31]-[32] Reducing $V_{G,offset}$ (typically of few tens of Volts) is necessary to interconnect GCS-based devices in series. This is because, once a device is driven into its resistive state by an applied $V_G > V_{G,offset}$, its output voltage ($V_{out}$) can be used to control another GCS-based device connected downstream, as long as $V_{out} \geq V_{G,offset}$ also for the latter device. A similar argument can be made regarding the interfacing of GCS-based logics to complementary metal-oxide semiconductor (CMOS) circuits (operating at voltages < 5 V)[33] to realise hybrid computing platforms with low energy dissipation – this remains one of the most promising potential applications of GCS.[25]

Another major challenge for applications is to achieve good reproducibility and scalability in GCS-based devices. To increase reproducibility, understanding the mechanism and the parameters behind GCS can be again crucial. For high scalability, top-down fabrication protocols based on subtractive patterning are preferable, since these are those adopted by semiconductor foundries to pattern CMOS circuits over large areas (≥ 6 inches).[34] Also, top-down protocols can help integration of GCS devices in S-based qubit platforms, where high-quality factor resonators are already made in the same way.[35]

In a recent study,[23] however, it has been shown that gated devices made following a top-down fabrication from Ss like Nb or NbTiN systematically do not exhibit GCS, unlike those made following a bottom-up approach (i.e., via additive patterning). The difference in the behaviour of these two types of devices has been ascribed to differences in the microstructural properties of the S nanoconstrictions, which have a rougher interface with the substrate and a more disordered surface (facing the gate electrode) in bottom-up fabricated devices.[23]

To investigate whether specific microstructural parameters and surface properties of the S nanoconstriction can lead to a systematic observation of GCS also in devices made with a top-down approach, in this work we study gated superconducting devices made by subtractive patterning from thin films of niobium rhenium (NbRe), for which GCS has not been yet explored. We have chosen NbRe as our S material because the results reported in ref. [23] suggest that disorder is an important parameter to observe GCS. Since our NbRe thin films are strongly disordered and consist of crystal grains (~ 1-2 nm) much smaller than the film thickness (~ 20 nm),[36] gated NbRe devices can help confirm the importance of disorder for GCS. We note that NbRe has other physical properties like high spin-orbit coupling (SOC), which also seems relevant for GCS,[25] and a non-centrosymmetric structure with an unconventional superconducting order parameter, at least in bulk single-crystal form.[37]-[38]

We find that our NbRe devices exhibit GCS, even though they are made using a top-down fabrication protocol. We also observe that GCS is only present when a specific gas mixture (consisting of Ar and $Cl_2$) is used for the fabrication, suggesting that the fabrication-induced modification of the S surface is crucial for the effect. GCS is instead systematically absent for devices etched with other gas mixtures, even in the presence of a significant $I_{leak}$ (> 10 nA). Last, in our NbRe devices showing GCS, the distance between the gate and the S nanoconstriction ($d_g$) is up to 300 nm, and therefore larger than that typically reported (< 100 nm) to observe GCS.[1],[8]-[9],[14]-[15]

We fabricate our gated NbRe devices from thin films using a top-down fabrication process. For all devices reported in this study (9 in total identified with labels from D1 to D9), we have used films of two different compositions and thicknesses, specifically 20-nm-thick $Nb_{0.18}Re_{0.82}$ (devices from D1 to D7) and 30-nm-thick $Nb_{0.10}Re_{0.90}$ (devices D8 and D9). All films have been deposited on $Al_2O_3$ substrates and are strongly disordered, as evidenced by X-ray diffraction analysis.[39]

Since we have followed a top-down fabrication process involving an etching step to make our devices, we refer to them as etched devices. As specified in the Methods section, the $Nb_{0.18}Re_{0.82}$ and $Nb_{0.10}Re_{0.90}$ films have been dry-etched using a negative resist and Al hard mask, respectively. Across devices, we have changed the gas mixture used for the etching process (Table S1), which we find to be the most crucial parameter for the GCS.

All our etched NbRe devices have been fabricated with a Dayem bridge geometry consisting of two large electrodes separated by a narrow constriction (bridge),[40] as shown in Figure 1a. The width of the bridge ranges between 50 nm and 80 nm, whilst the length is between 175 nm and 220 nm. In our devices, $d_g$ varies between 50 and 300 nm, with the gate electrode always placed only on one side of the S nanoconstriction (Figure 1b).

Figure 1c shows the resistance versus temperature, $R(T)$, for device D1, which has a superconducting critical temperature ($T_c$) ~ 6.1 K and a normal-state resistance ($R_N$) of ~ 1.1 kΩ. We define $T_c$ as the temperature ($T$) at which $R$ reaches 50% of $R_N$ at 10 K (Figure 1c). For comparison, the 20-nm-thick film used for the fabrication of this device has $T_c$ ~ 6.7 K, before patterning.[39] This suggests that our fabrication process in combination with the short superconducting coherence length ξ of NbRe (~ 5 nm; refs.[39]-[41]) preserves good superconducting properties in our Dayem bridges.

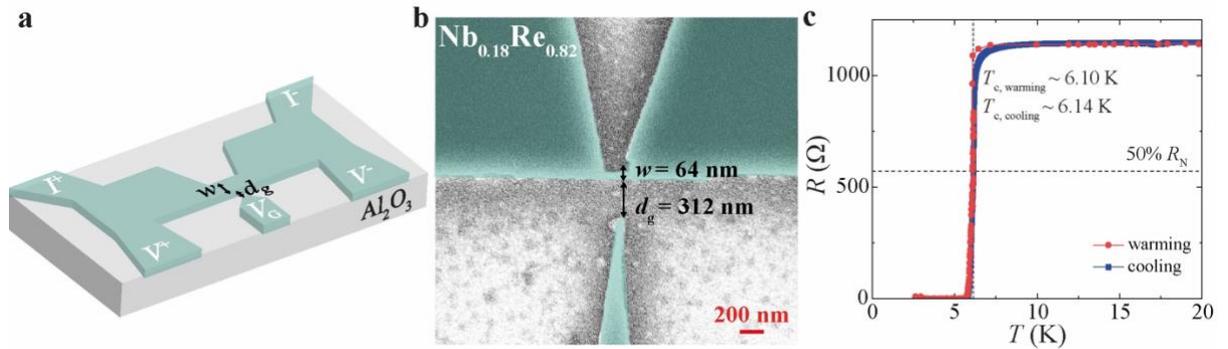

**Figure 1.** NbRe Dayem bridge devices. (a) Schematic of the geometry of a NbRe Dayem bridge device with configuration of the current ($I^-$ and $I^+$) and voltage ($V^-$ and $V^+$) pads and of the gate electrode ($V_G$), with false-colour scanning electron micrograph of a device (device D1) fabricated based on this layout in (b). (c) Resistance versus temperature $R(T)$ curve of the NbRe device D1 in (b) measured near its superconducting transition in both cooling (blue curve) and warming (red curve).

For all our devices, we have measured the evolution of $I_c$, extracted from current versus voltage $I(V)$ characteristics under an applied $V_G$. Our results discussed below suggest that only devices dry-etched using an Ar/Cl$_2$ gas mixture (D1 to D3) show GCS, whilst devices dry-etched with different gas mixtures (D4 to D9) show no GCS, even in the presence of a larger $I_{leak}$ and smaller $d_g$ compared to devices with GCS. We also observe a significant variation in the behaviour of the devices showing GCS, which suggests that different mechanisms may be at play across these devices (see below).

Figure 2b shows the $I_c(V_G)$ characteristics measured at different $T$s from 1.9 K to 5.8 K below $T_c$ (~ 6.1 K) for device D1, which reveals GCS with $V_{G,offset}$ ~ 55 V. As suggested in ref.[25], we define $V_{G,offset}$ as the point where the linear part of the $I_c(V_G)$ curve (i.e., the curve section where $I_c$ decays) intercepts the horizontal axis at $I_c = 0$. We also define $V_{G,onset}$ as the $V_G$ value of the $I_c(V_G)$ characteristic corresponding to a 10% drop in the $I_c$ measured at $V_G = 0$.[25] Compared to other devices reported in the literature and mostly made of elemental Ss (e.g., Nb, Al, Ti), for which $V_{G,offset}$ varies between 10 V and 40 V,[7],[9]-[10] $V_{G,offset}$ ~ 55 V measured for D1 is relatively large. The evolution of the $I_c(V_G)$ curves in an applied out-of-plane magnetic field $B_{ext}$ (Figure S1b) also confirms that the GCS effect persists until $B_{ext}$ suppresses superconductivity, consistently with previous studies.[1],[5],[16]

The observation of GCS in D1 and other devices (D2 to D3; see below) is remarkable for two reasons. First, $d_g$ for all these devices is three times larger than the typical $d_g$ (< 100 nm) necessary for GCS.[25] This also suggests that a $d_g$ reduction may lead to a further decrease in $V_{G,offset}$. Second, it has been reported that etched devices made of Ss different from NbRe (e.g., Nb or NbN) do not show GCS,[23] even when etched in the same gas mixture (Ar/Cl$_2$) for which we systematically observe GCS in NbRe devices.

Possible reasons why etched NbRe devices show GCS, unlike those in ref.[23], may be related to different physical properties of the Ss used. First, unlike Nb or NbN, NbRe is a non-centrosymmetric S. The non-centrosymmetric structure is linked also to strong SOC and to an unconventional superconducting order parameter.[37]-[38] Although we cannot quantify the role of these two physical properties on GCS – ad hoc theoretical investigations would be needed – our films are more disordered[35],[39] than the NbN and Nb thin films used in ref.[23]. The high disorder is not only evident from structural properties of the films, such as a grain size (~ 1-2 nm) much smaller than their thickness,[35] but also from their low-$T$ electronic transport properties. Indeed, the films used for the devices with GCS (D1÷D3) have a residual resistivity ratio (RRR) below 1 (Figure S2),[35]-[37] and their $T_c$ increases as resistivity $\rho$ gets larger (up to $\rho$ ~ 100 μΩ cm for a thickness of 20 nm)[35] – which are typical signatures of strong disorder in S thin films.[42]-[43] The films with composition Nb$_{0.10}$Re$_{0.9}$ (devices D8 and D9) have RRR >

1.5 and show no GCS, which further suggests that disorder plays an important role in our NbRe devices.

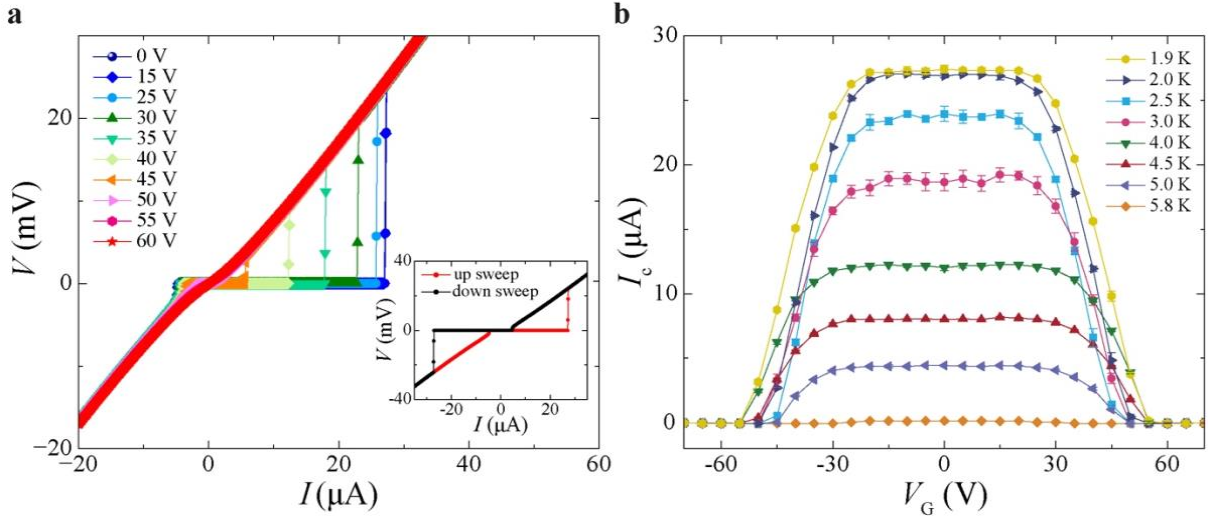

**Figure 2.** Gate-controlled supercurrent in NbRe device D1. (a) $I(V)$ characteristics measured at $T = 2.0$ K for a NbRe Dayem bridge (device D1) at different applied $V_G$ values (specified in the panel legend). The $I(V)$ curves shown in the main panel are measured whilst sweeping $I$ from negative to positive values (up-sweeping). The inset shows the $I(V)$ characteristic measured at the same $T = 2.0$ K and at $V_G = 0$ whilst up-sweeping (red curve) and down-sweeping (black curve) the bias current $I$. (b) $I_c(V_G)$ curves measured for the same device as in (a) at different $T$s (specified in the panel legend).

Across the NbRe devices showing GCS (D1 to D3), we observe a different $T$-dependence of $V_{G,\text{offset}}$ and of $I_{\text{leak}}$, from which we infer that the GCS is dominated by different physical mechanisms in these devices. In Figure S1, we report the $I_{\text{leak}}$ versus $V_G$, $I_{\text{leak}}(V_G)$, characteristics for device D1 measured at the same $T$s of the $I_c(V_G)$ curves in Figure 2b. At $V_{G,\text{onset}} \sim 24$ V, $I_{\text{leak}}$ is $\sim 100$ pA and is almost independent of $T$. On the other hand, $I_{\text{leak}}$ measured at $V_{G,\text{offset}}$ for the same device shows a strong $T$-variation, although $I_{\text{leak}}$ does not increase monotonically with increasing $T$. $V_{G,\text{offset}}$ also shows a similar $T$-dependence (Figure 2b).

For the other two devices with GCS (D2 and D3), $V_{G,\text{offset}}$ is of the same order of magnitude ($\sim 65$ V) as for device D1. For D2, however, $V_{G,\text{offset}}$ gets reduced monotonically, and simultaneously $I_{\text{leak}}$ (at $V_{G,\text{offset}}$) gets smaller as $T$ is increased (Figure S3). In device D2 therefore, GCS is mostly induced by $I_{\text{leak}}$ because, as $T$ is increased and superconductivity gets weaker, a lower $I_{\text{leak}}$ is measured whilst the device is driven into the normal state by $V_G$.

The $I_c(V_G)$ curves for device D3 in Figure 3a show that $V_{G,\text{onset}} \sim 40.8$ V and $V_{G,\text{offset}} \sim 64$ V are almost unaffected by $T$, which is opposite to what measured for device D2. The $I_{\text{leak}}$ values measured at $V_{G,\text{onset}}$ and $V_{G,\text{offset}}$ are also $T$-independent and equal to 0.2 nA and 3 nA, respectively.

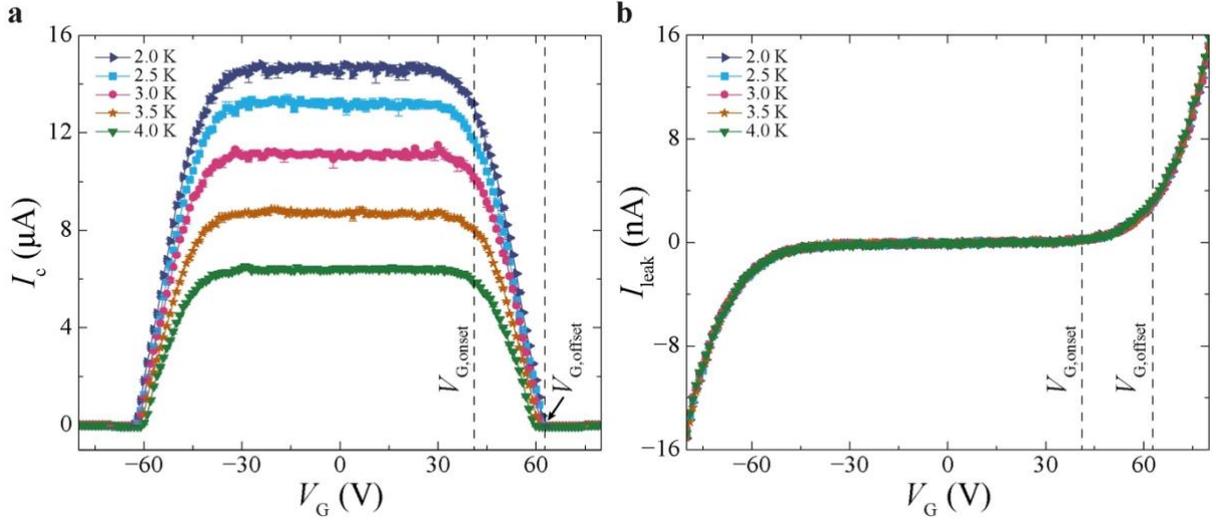

**Figure 3.** Gate-controlled supercurrent effect in NbRe device D3. (a) $I_c(V_G)$ curves measured for a NbRe Dayem bridge (device D3) at different $T$s (specified in the panel legend) with corresponding $I_{leak}(V_G)$ curves at the same $T$s in (b).

The different behaviour of the devices D1÷D3 is also shown in Figure 4, where we plot $V_{G,offset}$ as a function of $T$ for all these devices. The strong decrease of $V_{G,offset}$ with $T$ for device D2 again suggests that $I_{leak}$-induced Joule heating (scenario 2 above) may be the main mechanism responsible for the GCS in device D2, unlike for devices D1 and D3. The statement that Joule heating is likely not responsible for GCS in devices D1 and D3 seems to contradict what can be inferred from the comparison of the power dissipated by the gate $P_G = V_G \cdot I_{leak}$, with the power dissipated by Joule heating when the device switches to the resistive state – which we estimate as $P_N = R_N \cdot I_r^2$ ($I_r$ being the retrapping current). For device D3, for example, at $T = 2.0$ K, $P_G$ is ~ 8.2 nW at $V_{G,onset}$, where it is already comparable to $P_N$ ~ 16 nW ($R_N$ = 1.264 kΩ and $I_r$ = 3.54 µA at $V_G = 0$). This may suggest that Joule heating is the main contribution to GCS also for device D3.

However, the $I_{leak}$ measured for our devices (Figures 3, S1 and S3) and used to calculate $P_G$ does not correspond just to the $I_{leak}$ flowing through the S nanoconstriction, but it also includes contributions from the cryostat wiring. In our setup, $V_G$ is applied to the impedances of the gate electrode and of the setup cabling, which are connected in series between the $V_G$ generator and the setup electrical ground. As a result, we overestimate the actual $P_G$ dissipated by the Dayem bridge. To confirm this, we have also placed a reference resistor ($R_{ref}$) in series between the gate electrode and ground, to determine the actual $I_{leak}$ flowing across the bridge, based on the measurement of the voltage across $R_{ref}$. For this configuration, which we have tested on twin devices to devices from D1 to D3, we find $I_{leak}$ values lower by one order of magnitude than those measured for the total $I_{leak}$.

We are therefore confident that the actual $P_G$ dissipated within the NbRe bridges of devices D1÷D3 is lower than that calculated from the measured $I_{leak}$ by at least one order of magnitude. This estimate and the $T$-independent behaviour of $V_{G,offset}$, consistent with other reports where Joule heating has been ruled out,[1],[13],[22] suggest that different mechanisms are at play in devices D1 and D3.

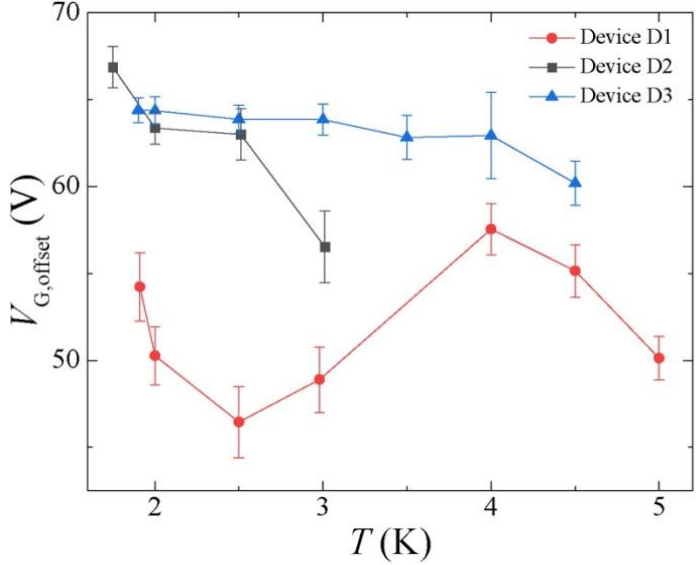

**Figure 4.** Comparison of etched NbRe devices. $V_{G,offset}(T)$ curves for D1 to D3 made by etching in Ar/Cl$_2$ gas mixture.

Another main result of our work is that the etching process, and specifically the gas mixture used, is crucial for the GCS observation. As reported in Table S1, in addition to devices etched in an Ar/Cl$_2$ mixture (D1 to D3), we have also made etched devices using Ar gas only (D4 and D5), Ar/SF$_6$ mixtures (D6 and D7) and devices etched with an Al hard mask, other than a negative resist, and an Ar/CF$_4$/O$_2$ mixture (D8 and D9). Nonetheless, none of the etched devices shows GCS up to $V_G \sim 100$ V (Figure S4), although some have a smaller $d_g$ (~ 50 nm) and larger $I_{leak}$ compared to devices D1 or D3. This suggests that, although $I_{leak}$ can play a role towards GCS, a large $I_{leak}$ *per se* is not sufficient to observe GCS in our etched NbRe devices.

Since in ref.[23] we have also used Ar/Cl$_2$ as etching gas for devices made of other Ss (e.g., Nb) and these showed no GCS, we conclude that intrinsic properties of NbRe like its high disorder in combination with surface properties, possibly activated by the etching gas, are key for the GCS observation. The importance of disorder in our NbRe films towards GCS can be verified, for example, by optimising the growth of NbRe thin films in single-crystalline form. To the best of our knowledge, however, the growth of NbRe in such form has not been achieved, possibly due to its large unit cell which makes epitaxial growth on lattice-matched substrates challenging.

Although the high structural disorder in our NbRe can assist GCS, disorder alone is not sufficient for GCS because devices etched with Ar or Ar/SF$_6$ show no GCS. The strong correlation between GCS and the etchant gas suggests that surface chemistry is instead crucial towards GCS.

It has been shown that the reaction of $Cl_2$ with Re results in the formation of $ReCl_5$ and other Re halides[44]-[45] with magnetic properties.[46]-[47] Since GCS is only observed in NbRe devices etched with $Cl_2$, these Re-based magnetic species forming on the surface may assist GCS, consistently with a recent theoretical proposal.[30] Future studies on etched devices made from Re thin films may further validate this argument.

In summary, we have shown that the NbRe surface properties, and how these are modified by the fabrication process, can lead to systematic GCS in etched NbRe devices.

We observe GCS despite an unusually-large $d_g$ (~300 nm), for which GCS is not observed in devices made of other Ss.[25] This suggests that other microstructural (e.g., high disorder) and/or physical properties (large SOC, unconventional order parameter) may support GCS in NbRe devices with the right surface properties.

Although an $I_{leak}$-induced mechanism can account for the GCS in our NbRe devices, our results show that this mechanism still has to find a S nanoconstriction with a suitable combination of surface and microstructural properties, to trigger GCS.

Our study therefore identifies a set of parameters that might be subject of further investigation, also of theoretical nature, aimed at understanding their role on the GCS. We also define a fabrication protocol that can be tested on other Ss similar to NbRe and that represents a first step towards achieving high reproducibility and scalability in GCS-based devices made with a top-down approach.

## ASSOCIATED CONTENT

The Supporting information contains: Experimental methods on fabrication and etching of NbRe thin films; $I_{leak}(V_G)$ curves at different $T$s and $I_c(V_G)$ curves at different $B_{ext}$ for device D1 (Figure S1); $R(T)$ curve for $Nb_{0.18}Re_{0.82}$ thin film from room $T$ down to low $T$ (Figure S2); $I_c(V_G)$ and $I_{leak}(V_G)$ curves at different $T$s for device D2 (Figure S3); $I_c(V_G)$ curves at two different $T$s for device D6 (Figure S4); Additional data on geometry and etching parameters for all devices (Table S1).

## AUTHOR INFORMATION

*Corresponding Authors*

**Elke Scheer** – Department of Physics, University of Konstanz, Konstanz 78457, Germany; https://orcid.org/0000-0003-3788-6979; Email: elke.scheer@uni-konstanz.de

**Angelo Di Bernardo** – Department of Physics, University of Konstanz, Konstanz 78457, Germany; Department of Physics, University of Salerno, Salerno *I*-84084, Italy; https://orcid.org/0000-0002-2912-2023; Email: adibernardo@unisa.it

*Authors*

**Jennifer Koch** - Department of Physics, University of Konstanz, Konstanz 78457, Germany; Email: Jennifer.koch@uni-konstanz.de

**Carla Cirillo** – CNR-Spin c/o University of Salerno, Salerno I-84084, Italy; https://orcid.org/0000-0001-8755-4484; Email: carla.cirillo@spin.cnr.it

**Sebastiano Battisti** – NEST, Istituto Nanoscienze-CNR and Scuola Normale Superiore, *I*-56-127 Pisa, Italy; https://orcid.org/0009-0006-2011-0196; Email: sebastiano.battisti@sns.it

**Leon Ruf** – Department of Physics, University of Konstanz, Konstanz 78457, Germany; https://orcid.org/0009-0005-3828-2669; Email: leon.ruf@uni-konstanz.de

**Zahra Makhdoumi Kakhaki** – Department of Physics, University of Salerno, Salerno *I*-84084, Italy; https://orcid.org/0000-0003-0330-8943; Email: zmahdoumikakhaki@unisa.it

**Alessandro Paghi** – NEST, Istituto Nanoscienze-CNR and Scuola Normale Superiore, *I*-56-127 Pisa, Italy; https://orcid.org/0000-0001-7452-7276; Email: alessandro.paghi@nano.cnr.it

**Armen Gulian** – Institute for Quantum Studies, Chapman University, Burtonsville 20866, MD, USA; https://orcid.org/0000-0002-4695-4059; Email: gulian@chapman.edu

**Serafim Teknowijoyo** – Institute for Quantum Studies, Chapman University, Burtonsville 20866, MD, USA; https://orcid.org/0000-0002-5542-5482; Email: serafimtw@gmail.com

**Giorgio De Simoni** – NEST, Istituto Nanoscienze-CNR and Scuola Normale Superiore, *I*-56-127 Pisa, Italy; https://orcid.org/0000-0002-2115-8013; Email: giorgio.desimoni@nano.cnr.it

**Francesco Giazotto** – NEST, Istituto Nanoscienze-CNR and Scuola Normale Superiore, *I*-56-127 Pisa, Italy; https://orcid.org/0000-0002-2115-8013; Email: francesco.giazotto@sns.it

**Carmine Attanasio** – Department of Physics, University of Salerno, Salerno *I*-84084, Italy; https://orcid.org/0000-0002-3848-9169; Email: cattanasio@unisa.it

*Author Contributions*

J. K. fabricated and measured the devices, with additional contributions and help from L. R., S. B., A. P., G. D. Most of the NbRe thin films were grown by Z. M. and C. C., with additional samples from S. T. and A. G. The measurement data were analyzed by J. K., E. S., A. D. B. with comments from C. A., F. G. and other authors. A. D. B. conceived the experiment and supervised it with E. S. The manuscript was written by A. D. B. and J. K. with inputs and comments from all authors.

*Notes*

The authors declare no competing financial interests.

## ACKNOWLEDGMENTS

We acknowledge support from the European Union's Horizon 2020 Research and Innovation Program under Grant Agreement No. 964398 (SuperGate). A. G. and S. T. also acknowledge support from the US ONR (grants No. N00014-21-1-2879, No. N00014-20-1-2442, and No. N00014-23-1-2866).